\documentclass[10pt,   showpacs, amsmath,amsfonts, amssymb,eqsecnum, twocolumn, draft, prd]{revtex4}%twocolumn, draft,

\usepackage{epsfig,eucal}

 \usepackage{amsfonts,amssymb}
\usepackage{amsmath}
\usepackage{graphicx}

\usepackage{bm}
\usepackage{color}
\definecolor{dark-green}{rgb}{0,0.7,0}
\definecolor{dark-blue}{rgb}{0,0.2,0.5}
\definecolor{med-blue}{rgb}{0,0.7,1}
\definecolor{mblue}{rgb}{0,0.2,1}
\definecolor{cnc}{rgb}{0.8,0,0}
\definecolor{light-red}{rgb}{1,0.8,0.8}
\definecolor{dark-yellow}{rgb}{1,0.8,0}
\definecolor{light-blue}{rgb}{0.8,0.9,1}
\definecolor{verylight-blue}{rgb}{0.93,0.95,1}
\definecolor{light-yellow}{rgb}{1,0.9,0.8}
\definecolor{grey}{gray}{0.88}
\definecolor{new-green}{rgb}{0.5,0.5,0.6}
\def\a{\alpha}
\def\b{\beta}
\def\g{\gamma}
\def\d{\delta}
\def\ep{\varepsilon}

\def\vp{\varphi}
\def\a{\alpha}
\def\b{\beta}

\def\g{\gamma}
\def\d{\delta}
\def\g{\gamma}

\begin{document}
 %\maketitle
%----------------------------
  \title{Photon propagator in skewon electrodynamics}
\author{Yakov Itin}
\email { itin@math.huji.ac.il}
%\\Institute of Mathematics, The Hebrew University of
% Jerusalem \\ and Jerusalem College of Technology, Jerusalem,
%  Israel. \\ email: {\tt itin@math.huji.ac.il}}

\affiliation{Institute of Mathematics, The Hebrew University of
  Jerusalem \\ and Jerusalem College of Technology, Jerusalem,
  Israel. }

%\affiliation{Institute of Mathematics, The Hebrew University of
%  Jerusalem \\ and Jerusalem College of Technology, Jerusalem,
%  Israel. \\ email: {\tt itin@math.huji.ac.il}}

 %---------------------------headings------------
%\pagestyle{myheadings}
%\markboth{Yakov Itin} {Yakov Itin %\qquad\qquad\qquad\qquad\qquad
%{Photon propagator in skewon electrodynamics}}

 %---------------------------headings------------
%\pagestyle{myheadings}
%\markboth{Yakov Itin} {Yakov Itin \qquad
%Skewon propagator}

%\markboth
%\begin{document}
%\maketitle
 \begin{abstract}
Electrodynamics with a local and linear constitutive law is used as a framework for models violating Lorentz covariance. The constitutive tensor of such a construction is irreducibly decomposed into three independent pieces. The principal part is the anisotropic generalisation of the standard electrodynamics. The two other parts, axion and skewon, represent non-classical modifications of electrodynamics. We derive the expression for the photon propagator in the Minkowski spacetime endowed with a skewon field. For a relatively small (antisymmetric) skewon field, a modified Coulom law is exhibited.
\end{abstract}
\pacs{75.50.Ee, 03.50.De, 46.05.+b, 14.80.Mz}
\keywords{Electrodynamics; Relativity; Constitutive law; 
Anisotropic media}
\date{\today}
\maketitle

%--------------------------------------------
\section{Introduction. Maxwell equations in linear response media }
%--------------------------------------------
 In Minkowski vacuum, the standard Maxwell electrodynamics is represented by a system of eight partial differential equations
\begin{equation}\label{intr1}
F_{[ij,k]}=0\,,\qquad {\rm and} \qquad F^{ij}{}_{,j}=J^i\, 
\end{equation}
 for six independent components of the antisymmetric tensor $F_{ij}$. 
Two  tensors $F^{ij}$ and  $F_{ij}$ appearing here are related by the use of the metric tensor
\begin{equation}\label{intr2}
 F^{ij}=\frac 12\, \left(g^{im}g^{jn}-g^{in}g^{jm}\right)F_{mn}\,.
\end{equation}
The Lorentz and CPT invariance of Eqs.(\ref{intr1} -- \ref{intr2})  is exhibited in the fact that propagation of   electromagnetic waves in this system generates the perfect light cone structure that is a fundamental expression of  Minkowski space-time geometry. 

Modern field models, such as loop  quantum gravity \cite{Gambini:1998it}, \cite{Bojowald:2004bb}, string theory  \cite{Kostelecky:1991ak}, \cite{Kostelecky:1995qk} and the very special relativity models \cite{Cohen:2006ky}, \cite{Gibbons:2007iu},  often predict violation of  the Lorentz and CPT invariance.  

The violation of the local Lorentz symmetry would also violates the Einstein equivalence principle,
In order to test the Einstein equivalence principle and to incorporate the Lorentz symmetry violation in electrodynamics.
one starts, see \cite{Ni:1977zz}, \cite{Ni6a}, \cite{Ni6b},\cite{Kostelecky:2002hh},   \cite{Kostelecky:2007fx}, with a generic extension of the electromagnetic Lagrangian
\begin{equation}\label{intr2a}
{\cal L}\sim \kappa^{ijkl}F_{ij}F_{kl}\,.
\end{equation}
 Due to its definition, the phenomenological tensor $\chi^{ijkl}$ is antisymmetric in two pairs of its indices and symmetric under the  exchange of these  pairs,  $\chi^{ijkl}=\chi^{klij}$. Altogether, it is left with 21 independent components. For most magnitudes  of these  parameters, the resulting field equations indeed have  solutions with breaking CPT symmetry and doubled light cone structure (birefringence). A similar approach was worked out in the wider context of standard model called standard model extension (SME), \cite{Kostelecky:2003fs}, and in gravity,  \cite{Bluhm:2007bd}. 
Even in a relatively simple case of  electrodynamics,  such construction is left with  a big set of undetermined  phenomenological parameters.  Moreover, a physical meaning of the individual terms of the Lagrangian (\ref{intr2a}) and of the corresponding phenomenological dimensionless coefficients remains undefined. 

In this paper, we apply an alternative approach based on a  modified constitutive relation instead of a modified Lagrangian. In other words, we are dealing with a non-trivial vacuum as a type of a medium and thus apply some constructions  familiar from solid state physics. 
In order to extend the vacuum system (\ref{intr1}--\ref{intr2})  to a {\it non-trivial medium}, it is  modified in the following way,  \cite{Post}, \cite{birkbook}:  
One considers two antisymmetric tensors -- the {\it field strength} $F_{ij}$ and the {\it excitation} $H^{ij}$. The field equations for this system are assumed to be of the standard Maxwell form:  
\begin{equation}\label{intr3}
F_{[ij,k]}=0\,,\qquad {\rm and} \qquad H^{ij}{}_{,j}=J^i\,. 
\end{equation}
Instead of the trivial relation (\ref{intr2}), a  generic linear local {\it constitutive relation} 
 between two tensor fields,  
\begin{equation}\label{intr4}
H^{ij}=\frac 12\, \chi^{ijmn}F_{mn}\,, 
\end{equation}
is postulated. Since the metric tensor is not involved  in   Eqs.(\ref{intr3}--\ref{intr4}), this approach  is often referred to as {\it premetric electrodynamics}. 
The construction presented by Eqs.(\ref{intr3}--\ref{intr4}) allows to describe different types of equivalence principle violation models \cite{Ni:1977zz}, and Lorentz violation models in electrodynamics \cite{Kostelecky:2002hh}, \cite{Kostelecky:2007fx}. It is wide enough to include in a natural way the vacuum birefringence \cite{Lammerzahl:2004ww}, \cite{Itin:2005iv}, the axion construction \cite{Ni:1977zz}, \cite{Carroll:1989vb},   \cite{Itin:2004za}, \cite{Itin:2007wz}, \cite{Itin:2007cv}  and some special types of Finsler geometry  \cite{Pfeifer:2011tk},  \cite{Itin:2014uia}. Recently an intensive theoretical study of wave propagation in the system (\ref{intr3}--\ref{intr4}) was provided, see 
\cite{Obukhov:2000nw}, \cite{Rubilar:2007qm}, \cite{L-S}, \cite{Itin:2009aa}, \cite{Perlick:2010ya}, \cite{Favaro:2010ys}, \cite{Dahl:2011he}. An important output of this analysis is a proof that in the  linear response system (\ref{intr3}--\ref{intr4}), the generalized Fresnel hypersurface (dispersion relation) is quartic in general.  Thus birefringence turns out to be a generic property of these models.

In (\ref{intr4}), the {\it constitutive tensor} $\chi^{ijmn}$ is not restricted by the pair interchanging symmetry, i.e.,  $\chi^{ijmn}\ne \chi^{mnij}$, in general. Consequently,  $\chi^{ijmn}$ has 36 independent components, i.e, 15 extra components relative to the Kostelecky tensor $\kappa^{ijmn}$.  This additional set of components forms a $GL(4)$-irreducible part of the constitutive tensor, called {\it skewon}, \cite{birkbook}. It is a main object of our current consideration. 

In the pioneer work of Obukhov and Hehl \cite{Obukhov:2004zz}, various types of  the skewon media with the corresponding dispersion relations were  outlined. 
In \cite{Itin:2014gba}, we presented a new form of dispersion relation in the skewon sector of the premetric electrodynamics and derive some new features of this model, such as {\it superluminal wave propagation}  and {\it Higgs-type symmetry breaking}. 
The analysis given in \cite{Obukhov:2004zz}, \cite{Itin:2014gba}, \cite{Itin:2013ica},  \cite{Ni:2013uwa},  and \cite{Itin:2014rwa} was restricted to the wave propagation problems, i.e., to the vacuum case. 
In the present paper, we extend  this analysis to the non-empty regions and derive the explicit expression of the skewon modified photon propagator (Green's function in the momentum representation). This tensor is important for a description of the interaction between waves and charges and for an account of the quantum effects. 
Usually, the propagator tensor in Lotents violation models is derived from the Lagrangian, see \cite{Casana:2009xs},\cite{Schreck:2011ai}, \cite{Lingli:2011ei}. Since skewon represents the non-Lagrangian sector of the model, this method is not available in our case. Instead, we use an algebraic technique based on the notion of the second adjoint tensor, see \cite{Itin:2007cv}, \cite{Itin:2007av}.  
The skewon propagator is shown to have an ordinary form of the second order tensor that is singular on the doubled lightcone. Due to the anisotropic property of the skewon field, the propagator has a non zero antisymmetric part. We derive the expressions in the Feynman, Landau and Coulomb gauge. 
We apply the latter type of propagator for derivation of the skewon modification of the Coulomb law. The ellipsoidal equipotential surfaces in this case represent the violation of the isotropy in antisymmetric skewon model.   

The organization of the paper is as follows. In the next section, we present the geometric optic approximation of the model (\ref{intr3}--\ref{intr4}). As a result,  we arrive in a system of algebraic equations for the electromagnetic potential. The coefficients of this system are linear in the constitutive tensor components and quadratic in the wave covector. On the tangent bundle, these coefficients compose  a second order tensor that we call  {\it optic tensor}.  
In section 3  we define the photon propagator and derive its generic expression for the case of Maxwell's  system with a generic linear constitutive relation defined on a metric space. 
In section 4, we derive the explicit expression of the photon propagator for the case of a medium with a simplest Minkowski  principal part but a most generic skewon part. 
Application of this formalism for  specific examples of antisymmetric skewon  is presented in section 5. Here we derive a specific example of the skewon modified Coulomb law. 
In the conclusion section, we  discuss some extensions of our construction. 
Certain relevant algebraic facts and some explicit  matrix calculations are given in Appendix. 
%%%%%%%%%%%%%%%%%%%%%%%
\section{Characteristic system and dispersion relation}
In order to discuss the solutions of the system (\ref{intr3})--(\ref{intr4}), we start with  the ansatz 
\begin{equation}\label{char1}  
F_{ij}=(1/2) \left(A_{i,j}-A_{j,i}\right)\,.
 \end{equation}
that solves the homogeneous  equation in  terms of  vector potential. 
Substituting (\ref{intr4}) and (\ref{char1})   into the second equation of (\ref{intr3}),  we have 
\begin{equation}\label{char2}  
 \left(\chi^{ijmn}A_{m,n}\right)_{,j}=J^i\,.
 \end{equation}
Now we apply the {\it geometric optics approximation}. First we assume the tensor  $\chi^{ijmn}$ to be a slowly varying function of the coordinates  with respect to the fast   varying function $A_m$. Consequently, Eq.(\ref{char2}) is approximated as  
\begin{equation}\label{char3}  
\chi^{ijmn}A_{m,nj}=J^i\,.
 \end{equation}
Using the Fourie transforms of the current 
\begin{equation}\label{char4} 
J^i(x)=\int j^i(q) \exp\left({iq_kx^k}\right)d^4q\,,
 \end{equation}
and of the potential
\begin{equation}\label{char4a} 
A_m(x)=\int a_m(q)\exp\left({iq_kx^k}\right)d^4q\,,
 \end{equation}
 we obtain from Eq.(\ref{char3}) the {\it characteristic system} 
\begin{equation}\label{char5} 
 \big(\chi^{ijmn}q_jq_n\big)a_m=j^i\,.
 \end{equation}
In the following, it is convenient to use the notation 
\begin{equation}\label{char6} 
M^{im}= \chi^{ijmn}q_jq_n\,.
 \end{equation}
We refer to $M^{ij}$ as the {\it optic tensor}. It is   analogous to the acoustic (Christoffel) tensor of  elasticity theory. 
Consequently, we rewrite   Eq.(\ref{char5}) in the form of a general  system of four linear equations for four independent variables
\begin{equation}\label{char7} 
M^{im}a_m=j^i\,.
 \end{equation}
The basic feature of this system is that the matrix $M$ is singular and thus for given functions $M^{im}(q)$ and $j^i(q)$, the solution $a_m$  to Eq.(\ref{char7}) is not unique. 
Indeed, we immediately  observe from Eq.(\ref{char6}) two   linear relations
\begin{equation}\label{char13} 
M^{ij}q_j=0\,, \qquad {\rm and}\qquad M^{ij}q_i=0\,.
 \end{equation}
Being  trivial consequences of the symmetries of $\chi^{ijmn}$, these relations nevertheless reflect the basic physical facts of electromagnetism:
\begin{itemize}
\item[(i)] The first relation of Eq.(\ref{char13}) expresses the {\it gauge invariance} of our system: If  $a_m$ is a solution of Eq.(\ref{char7}) then  $a_m+Cq_m$ is also a solution. This symmetry  is an algebraic expression of the gauge invariance of the electromagnetic potential $A_i\to A_i+\partial_i\vp$. 
\item[(ii)]The second relation of  Eq.(\ref {char13}) corresponds to the {\it electric charge conservation law.} Indeed, the multiplication of both sides of Eq.(\ref{char7}) by $q_i$ yields $j^iq_i=0$. It is an algebraic expression of the  charge conservation law $J^i{}_{,i}=0$. 
\end{itemize}
Now we use the following algebraic fact (see \cite{Itin:2009aa} for the proof): 
For a tensor $M^{ij}$ satisfying the relations  (\ref{char13}),  
the adjoint tensor $A_{ij}={\rm adj}\, M$ is proportional to the tensor square of the covector $q_i$, i.e., 
 \begin{equation}\label{prop8} 
A_{ij}=\lambda(q)q_iq_j\,.
 \end{equation}
Here $\lambda(q)$ is a fourth order homogeneous polynomial of the wave covector  $q_i$. We recall that a polynomial is called homogeneous of the order $r$ if for every constant $C$ it satisfies $\lambda(Cq)=C^r\lambda(q)$. 
Since the adjoint matrix is constructed from the determinants of $3\times 3$ matrices, $\lambda(q)$ is a third-order homogeneous polynomial of the constitutive tensor $\chi$. 

In   source-free regions where $J^i=0$, a non-trivial solutions to Eq.(\ref{char7}) exist only if $A_{ij}=0$. Consequently, Eq.(\ref{prop8}) yields the {\it dispersion relation} for the electromagnetic waves in the form
 \begin{equation}\label{prop9} 
\lambda(q)=0\,.
 \end{equation}
In the Minkowski case, Eq.(\ref{prop9}) reads  $\left(g^{ij}q_iq_j\right)^2=0$. It yields the standard Lorentz light cone $g^{ij}q_iq_j=0$. 
%----------------------------------------
\section{Propagator for linear response media}
In this section, we study  solutions of the system (\ref{char7}) in a region where $J^i\ne 0$. These solutions are represented by the Green functions (Green tensors).  We use the momentum-space representation similar to the usually used in quantum field theory   and  refer  to Green functions as {\it photon  propagators}. 
%%%
\subsection{The propagator tensor and its gauge transformations}
 A formal solution of the linear system  in  Eq.(\ref{char7}) can be  written  as 
\begin{equation}\label{prop1} 
a_i=D_{ik}j^k\,.
 \end{equation}
The tensor $D_{ik}$ is called the {\it  propagator tensor}. 
 In standard quantum electrodynamics based on the Minkowski constitutive tensor, it  suffices  to consider  symmetric propagators, $D_{ik}=D_{ki}$. We will see, however, that in the case of a generic linear constitutive relation, the propagator tensor is asymmetric. 
 Notice that the antisymmetric part of the photon propagator is already derived in the axion model \cite{Itin:2007cv}.

Any  second order tensor can be decomposed into symmetric and antisymmetric parts:  
 \begin{equation}\label{prop2-1} 
D_{ij}=\check{D}_{ij}+\tilde{D}_{ij}\,.
 \end{equation}
Here  the  symmetric  part is denoted by 
 \begin{equation}\label{prop2-1a} 
\check{D}_{ij}=D_{(ij)}= (1/2)\left(D_{ij}+D_{ji}\right) \,,
 \end{equation}
%$\check{D}_{ij}=D_{(ij)}$ 
while the antisymmetric part is denoted by 
 \begin{equation}\label{prop2-1b} 
\check{D}_{ij}=D_{[ij]}=(1/2)\left(D_{ij}-D_{ji}\right) \,.
 \end{equation}
%$\tilde{D}_{ij}=D_{[ij]}$, i.e., 
Evidently, 
\begin{equation}\label{prop2-4} 
\check{D}_{ij}=\check{D}_{ji}\,,\qquad \tilde{D}_{ij}=-\tilde{D}_{ji}\,.
 \end{equation}
In  standard electrodynamics, propagator has only the simplest symmetric scalar part ${D}_{ij}\sim D g_{ij}$.  Since the diagonal Minkowski metric is considered, it means  the vectors $a_i$ and $j_i$ are proportional one to another. In particular, static charge creates an electric field only, and no magnetic.
Additional terms in the propagator tensor must exhibit  a  richer class of phenomena.

%\subsection{Gauge transformations of the  propagator}
 Due to the electric charge conservation, $q_kj^k=0$, the propagator $D_{ik}$ in Eq.(\ref{prop1}) is defined only up to an  additional  term proportional to  $q_k$. Also a term proportional to $q_i$ can be freely added to $D_{ik}$. It is due to the gauge invariance of the potential,  $a_i\to a_i+Cq_i$. 
Consequently, we arrive  to the generic {\it gauge transformation of the propagator} 
 \begin{equation}\label{prop2} 
D_{ik}\to D_{ik}+q_i\phi_k+q_k\psi_i \,.
 \end{equation}
Here $\phi_k$ and $\psi_i$ are arbitrary functions of the wave covector $q_i$. We  recall that for an ordinary symmetric propagator, the transformation (\ref{prop2}) is used with only one arbitrary function, i.e., the relation $\phi_i=\psi_i$ is assumed.   When $\phi_k$ and $\psi_i$ are treated as covectors, we obtain in (\ref{prop2})  a Lorentz invariant gauge transformation.  Alternatively, in some situations, it may be advantageous to use a non-invariant Coulomb-type gauge. In this case, some components of $\phi_k$ and $\psi_i$ are treated as fixed or even equal to zero.

We write the gauge transformation laws for the symmetric and antisymmetric parts  separately  
 \begin{equation}\label{prop2a} 
\check{D}_{ik}\to \check{D}_{ik}+q_i\rho_k+q_k\rho_i \,,
 \end{equation}
and 
\begin{equation}\label{prop2b} 
\tilde{D}_{ik}\to \tilde{D}_{ik}+q_i\sigma_k-q_k\sigma_i \,.
 \end{equation}
 Here,  
\begin{equation}\label{prop2c}
\rho_k=(1/2)(\phi_k+\psi_k),\quad  {\rm {and}} \quad\sigma_k=(1/2)(\phi_k-\psi_k)\,. 
 \end{equation}

\subsection{Feynman, Landau and Coulomb propagators}
The transformation relations (\ref{prop2}) may be used for simplification the expressions  of the propagator.

\subsubsection{Feynman propagator} 
Starting with an arbitrary propagator $D_{ik}$ and choosing certain  special functions for $\psi_k$ and $\phi_i$,  we can freely remove from $D_{ik}$ all the terms proportional to $q_i$ and $q_k$. 
In a correspondence with the usual practice, we will call such simplest  propagator, without  longitudinal components, as  the {\it Feynman propagator} and denote it as ${}^F\!{D}_{ik}$. We denote by ${}^F\!\check{D}_{ik}$  its  symmetric part, and by  ${}^F\!\tilde{D}_{ik}$ its  remaining antisymmetric part. 
\subsubsection{Landau propagator} 
The Landau propagator is defined to be purely transverse, i.e., $ {D}_{ik}q^k=0$. Notice that Feynman's and Landau's propagators are both covariant. 
In asymmetric case, the generalization of the {\it Landau-gauge  propagator} ${}^L\!{D}_{ik}$  is naturally to define by two independent conditions 
 \begin{equation}\label{prop2cx} 
{}^L\!{D}_{ik}q^k=0\,,\qquad {}^L\!{D}_{ik}q^i=0 \,. 
 \end{equation}
These equations are equivalent to two separate conditions for symmetric and antisymmetric parts  respectively
 \begin{equation}\label{prop2cc} 
{}^L\!\check{D}_{ik}q^k=0\,,\qquad {}^L\!\tilde{D}_{ik}q^k=0 \,.
 \end{equation}
The Landau  propagator can be derived by the gauge transformation from the Feynman propagator. 
For the purposes of the derivation, it is convenient to deal with the symmetric part ${}^L\!\check{D}_{ik}$ and antisymmetric part ${}^L\!\tilde{D}_{ik}$  separately. 

For the symmetric part of the Landau  propagator, we write
 \begin{equation}\label{prop2add1} 
{}^L\!\check{D}_{ik}={}^F \!\check{D}_{ik} +q_i\phi_k+q_k\phi_j \,.
 \end{equation}
We use the first gauge condition of (\ref{prop2cc}) to get 
  \begin{equation}\label{prop2add2} 
{}^F \!\check{D}_{ik}q^k+q_i(q,\phi)+q^2\phi_i =0\,.
 \end{equation}
Multiplying two sides of this equation by $q^i$, we obtain   
 \begin{equation}\label{prop2add3} 
(q,\phi) =-\frac\Delta{2q^2}\,,\qquad {\rm where}\qquad 
\Delta={}^F\!{D}_{mn}q^mq^n\,.
 \end{equation}
Consequently,  Eq. (\ref{prop2add2}) yields
 \begin{equation}\label{prop2add4} 
\phi_i=-{}^F \!\check{D}_{ik}\,\frac {q^k}{q^2} +\frac\Delta{2q^4}q_i\,.
 \end{equation}
From Eq.(\ref{prop2add1}), the symmetric part of the Landau propagator is expressed now as   
 \begin{equation}\label{prop2add5} 
{}^L\!\check{D}_{ik}={}^F \!\check{D}_{ik} - \left(
{}^F \!\check{D}_{im}q_k+{}^F \!\check{D}_{km}q_i\right) \frac{q^m}{q^2}+\frac\Delta{q^4}\,q_iq_k\,.
 \end{equation}

For the antisymmetric part of the propagator, we write
 \begin{equation}\label{prop2add6} 
{}^L\!\tilde{D}_{ik}={}^F \!\tilde{D}_{ik} +q_i\psi_k-q_k\psi_i \,.
 \end{equation}
Contracting with $q^k$, we obtain
 \begin{equation}\label{prop2add7x} 
{}^L\!\tilde{D}_{ik}q^k={}^F \!\tilde{D}_{ik}q^k +q_i(\psi,q) -q^2\psi_i =0\,.
 \end{equation}
This algebraic equation for $\psi$ yields  $(\psi,q)=0$. 
Consequently, we obtain now 
 \begin{equation}\label{prop2add7} 
\psi_i={}^F \!\tilde{D}_{ik}\,\frac {q^k}{q^2}\,.
 \end{equation}
Hence, the relation between the antisymmetric parts of the Feynman and Landau propagators is given by 
 \begin{equation}\label{prop2add8} 
{}^L\!\tilde{D}_{ik}={}^F \!\tilde{D}_{ik}+\left({}^F \!\tilde{D}_{km}q_i-
{}^F \!\tilde{D}_{im}q_k\right)\frac{q^m}{q^2}\,.
 \end{equation}
The sum of the right hand sides of Eqs.(\ref{prop2add5}, \ref{prop2add8}) yields finally the expression of the  Landau propagator in term of the Feynman one 
 \begin{equation}\label{prop2add9} 
{}^L\!{D}_{ik}={}^F \!{D}_{ik}-\left({}^F \!{D}_{im}q_k+
{}^F \!{D}_{mk}q_i\right)\frac{q^m}{q^2}+\frac\Delta{q^4}\,q_iq_k\,.
 \end{equation}
It is straightforward to check for this expression the identities ${}^L\!{D}_{ik}q^i=0$ and ${}^L\!{D}_{ik}q^k=0$. 

\subsubsection{Coulomb propagator} 
In  static electromagnetic problems, it is useful to use a propagator that is covariant only under the group $SO(3)$ of  3-dimensional rotations. We denote such Coulomb-gauge propagator as ${}^C\!{D}_{ij}$. 
We use the $(1+3)$-decomposition of the wave vector as 
 \begin{equation}\label{col1}
q_i=(\omega, k_\a)\,, \qquad  q^i=(-\omega, k^\a)\,.
 \end{equation}
Here and in the sequel, the Greek indices take values  in the range $\{1,2,3\}.$  We use the standard Minkowski metric in the Cartesian coordinates $g^{ij}={\rm diag}  (-1,1,1,1)$. 
In the case of an asymmetric propagator, it is natural to extend the standard Coulomb gauge condition, ${}^C\!{D}_{i\mu}k^\mu=0$,  to a pair of two separate  conditions for symmetric and skew-symmetric parts respectively,
 \begin{equation}\label{col2}
{}^C\!{D}_{i\mu}k^\mu=0\,,\qquad {}^C\!{D}_{\mu i}k^\mu=0\,,
 \end{equation}
or, equivalently, 
 \begin{equation}\label{col2a}
{}^C\!\check{D}_{i\mu}k^\mu=0\,,\qquad {}^C\!\tilde{D}_{i\mu}k^\mu=0\,.
 \end{equation}
Let us start with the symmetric part, ${}^C\!\check{D}_{i\mu}$. Using the transformation law 
(\ref{prop2a}), we write
 \begin{equation}\label{prop2ax} 
{}^C\!\check{D}_{ik}= {}^F\!\check{D}_{ik} +q_i\rho_k+q_k\rho_i \,,
 \end{equation}
The first equation of (\ref{col2a}) yields
 \begin{equation}\label{col3}
{}^C\!\check{D}_{i\mu}k^\mu={}^F\!\check{D}_{i\mu}k^\mu+q_i(k,\rho) +\rho_ik^2=0\,.
 \end{equation}
Here we use  three dimensional notations: $k^2=k_\a k^\a$ and $(k,\rho)=k^\a \rho_\a$. 
 For $i=0$, we have  in Eq.(\ref{col3}) 
\begin{equation}\label{col4}
{}^F\!\check{D}_{0\mu}k^\mu+\omega(k,\rho) +\rho_0k^2=0\,,
 \end{equation}
and, for $i=\a$,
\begin{equation}\label{col5}
{}^F\!\check{D}_{\a\mu}k^\mu+k_\a(k,\rho) +\rho_\a k^2=0\,.
 \end{equation}
Multiplying the latter equation by $k^\a$ we derive
 \begin{equation}\label{col6}
(k,\rho) =-\frac{\cal D}{2k^2},\qquad{\rm where}\qquad {\cal D}= {}^F\!D_{\a\b}k^\a k^\b\,.
 \end{equation}
Consequently, Eqs.(\ref{col4},\ref{col5}) yield 
 \begin{equation}\label{col7}
\rho_0=\frac{\omega\cal D}{2k^4}- \frac {{}^F\!\check{D}_{0\mu}k^\mu}{k^2}\,,
 \end{equation}
and
 \begin{equation}\label{col8}
\rho_\a=\frac{k_\a\cal D}{2k^4}- \frac {{}^F\!\check{D}_{\a\b}k^\b}{k^2}\,.
 \end{equation}
Substituting (\ref{col7}) and (\ref{col8}) into Eq.(\ref{prop2a}) we obtain for the symmetric part of the Coulomb propagator
 \begin{equation}\label{col9}
{}^C\!{D}_{00}=%{}^F\!{D}_{00}+2\omega \rho_0=
{}^F\!{D}_{00}+\frac{\omega^2\cal D}{k^4}- 2\frac {{}^F\!\check{D}_{0\mu}\omega k^\mu}{k^2}\,,
 \end{equation}
\begin{equation}\label{col10}
{}^C\!\check{D}_{0\b}=
{}^F\!\check{D}_{0\b}+\frac{\omega k_\b\cal D}{k^4}- 
\frac { k^\mu}{k^2} \left({}^F\!\check{D}_{\b\mu}\omega + {}^F\!\check{D}_{0\mu}k_\b\right)\,,
 \end{equation}
and
\begin{equation}\label{col11}
{}^C\!\check{D}_{\a\b}=
{}^F\!\check{D}_{\a\b}+\frac{ k_\a k_\b\cal D}{k^4}- 
\frac { k^\mu}{k^2} \left({}^F\!\check{D}_{\a\mu}k_\b + {}^F\!\check{D}_{\b\mu}k_\a\right)\,.
 \end{equation}

We consider now the skew-symmetric Coulomb propagator. Due to
the transformation law 
(\ref{prop2a}), it can be written as 
 \begin{equation}\label{col12} 
{}^C\!\tilde{D}_{ik}= {}^F\!\tilde{D}_{ik}+q_i\sigma_k-q_k\sigma_i \,,
 \end{equation}
In $(1+3)$-notations, it reads
 \begin{equation}\label{col13} 
{}^C\!\tilde{D}_{0\a}= {}^F\!\tilde{D}_{0\a} +\omega\sigma_\a-k_\a\sigma_0 \,,
 \end{equation}
and
 \begin{equation}\label{col13a} 
{}^C\!\tilde{D}_{\a\b}= {}^F\!\tilde{D}_{\a\b} +k_\a\sigma_\b-k_\b\sigma_\a \,.
 \end{equation}
%Applying the Coulomb gauge condition we get from the  latter equation 
% \begin{equation}\label{col14} 
%{}^C\!\tilde{D}_{\a\b}k^\b= {}^F\!\tilde{D}_{\a\b}k^\b +k_\a(\sigma, k)-k^2\sigma_\a=0 \,.
% \end{equation}
The solution of the Coulomb gauge condition $\tilde{D}_{i\mu}k^\mu=0$ with the expressions (\ref{col13}) and (\ref{col13a}) substituted is not unique (a term proportional to $k_\a$ can be freely added to $\sigma_\a$). But, for our purposes,  we need only a partial solution. We observe that 
 \begin{equation}\label{col14} 
\sigma_0=\frac{{}^F\!\tilde{D}_{0\b}k^\b}{k^2}\,,\qquad \sigma_\a=\frac{{}^F\!\tilde{D}_{\a\b}k^\b}{k^2} \,.
 \end{equation}
is one of such solutions. Consequently we get the skew-symmetric part of the Coulomb propagator
\begin{equation}\label{col15} 
{}^C\!\tilde{D}_{0\b}= {}^F\!\tilde{D}_{0\b} +\frac {k^\mu}{k^2}\left({}^F\!\tilde{D}_{\b\mu}  \omega-{}^F\!\tilde{D}_{0\mu}k_\b \right) 
 \,,
 \end{equation}
and
 \begin{equation}\label{col15a} 
{}^C\!\tilde{D}_{\a\b}= {}^F\!\tilde{D}_{\a\b} +\frac {k^\mu}{k^2}\left({}^F\!\tilde{D}_{\b\mu}k_\a -{}^F\!\tilde{D}_{\a\mu}k_\b \right) \,.
 \end{equation}

\subsection{Propagator of a generic linear response system}
In order to derive the expression for  $D_{ik}$ for the singular system (\ref{char7})  we will use the {\it second adjoint} of the matrix $M^{ij}$.  It comes instead of the ordinary first adjoint (a factor of the inverse matrix) used for the regular systems. Please note that the adjoint tensors can be defined for any tensor and this concept does not require a metric structure.  In Appendix A, we provide formal definitions and basic formulas related to the first and the second adjoint tensors. Recall that the first adjoint of a $(2,0)$-order tensor $M^{ij}$ is expressed by a $(0,2)$-order tensor $A_{ij}$. The full contraction of these two tensors  is proportional to the determinant of the matrix,  $M^{ij}A_{jk}\sim{\rm det\,}M$. The second adjoint of a $(2,0)$-order tensor $M^{ij}$ is expressed by a $(0,4)$-order tensor $B_{ijkl}$. Its contraction with the tensor  $M^{ij}$ is proportional to the first adjoint $A_{ij}$, see (\ref{ap-prop6}). 

Our goal is to derive the solution $a_m$ of  the singular system  (\ref{char7}). 
It is instructive to recall how one  deals with  a similar regular  system $M^{ik}a_k=j^i$ in  basic linear algebra. 
First one multiplies both sides of  this equation by the adjoint matrix $A_{ki}={\rm adj}(M)$. 
Due to the Laplace identity (\ref{det3}), one remains  with the equation 
$
\left({\rm det} \,M\right)a_i=A_{ik}j^k\,
$
For the invertible matrix $M$, the unique solution is obtained by dividing both sides of the equation by the scalar factor ${\rm det} \,M$, i.e. by forming the inverse matrix. 
 In our case, the  above procedure does not work because the matrix is not invertible, i.e., ${\rm det} \, M=0$. 

For a singular system, we apply a similar procedure but with the use of the second adjoint $B_{ijkl}$ instead of the first adjoint $A_{ij}$. We multiply  two sides of the equation 
\begin{equation}\label{prop5a} 
M^{ij}a_j=j^i
 \end{equation}
  by  $B_{irks}$ to obtain
 \begin{equation}\label{prop6} 
B_{irks}M^{ij}a_j=B_{irks}j^i\,.
 \end{equation}
Applying the second order Laplace expansion (\ref{ap-prop6}), we rewrite  this equation as 
 \begin{equation}\label{prop7x} 
A_{rs}a_k-A_{rk}a_s=B_{irks}j^i\,.
 \end{equation}
Substituting here the expression for the adjoint (\ref{prop8}), we obtain
\begin{equation}\label{prop8x} 
\lambda q_r(q_s a_k-q_ka_s)=B_{irks}j^i\,.
 \end{equation}
The problem now is to extract the potential $a_i$ from this equation. 
In \cite{Itin:2007av}, we provide a metric-free procedure  based 
on the homogeneity of our algebraic equation. In the presence of the metric structure,  the problem can be solved much simpler. 
First we multiply both sides of Eq.(\ref{prop8x}) by the metric tensor $g^{rs}$ and use the Lorenz gauge condition $a_sq^s=0$. 
Consequently,  Eq.(\ref{prop8x}) takes the form 
 \begin{equation}\label{prop9x} 
\lambda q_rq_ja_s=-B_{ijrs}j^i\,.
 \end{equation}
Contracting now two sides of this equation with the use of the metric tensor, we remain with 
 \begin{equation}\label{prop10} 
a_k=\frac 1{\lambda q^2} \, g^{rs} B_{irks}j^i\,.
 \end{equation}
This expression can be made to satisfy the  Lorenz gauge condition by adding a term of the form $Cq_k$ with a specially chosen scalar $C$. Since we are interested in a generic propagator and remember the propagator gauge freedom  (\ref{prop2}), we do not need to redefine the potential. 
It is instructive to check our solution (\ref{prop10}) by substituting it into the original equation (\ref{prop5a}). 
We calculate 
\begin{equation}
M^{pk}a_k= \frac 1{\lambda q^2} \, g^{rs} M^{pk}B_{irks}j^i\,.
 \end{equation}
Using the rule (\ref{ap-prop6}), we obtain
\begin{equation}
M^{pk}a_k= \frac 1{\lambda q^2} \, g^{rs} \left(\d^p_iA_{rs}-\d^p_rA_{is}\right)j^i\,.
 \end{equation}
Substituting here the expression for the adjoint (\ref{prop8}), we get
\begin{equation}\label{prop11aa}
M^{pk}a_k= \frac 1{ q^2} \, g^{rs} \left(\d^p_iq_rq_s-\d^p_rq_iq_s\right)j^i\,.
 \end{equation}
Using the charge conservation equation $j^kq_k=0$, we obtain in the right hand side of Eq.(\ref{prop11aa}) the $p$-th component of the current  $j^p$. 
Thus  (\ref{prop10}) indeed represents a solution to Eq.(\ref{prop5a}). 
Even if it is only a special solution, the whole set of solutions can be readily  reinstated by the gauge transformation of the potential.

From Eq.(\ref{prop10}), we extract the propagator tensor 
 \begin{equation}\label{prop11} 
{}^F\!D_{ij}=\frac 1{\lambda q^2} \,  B^{m}{}_{ijm}\,.
 \end{equation}
This expression is derived by removing the longitudinal terms, thus it must be treated as the  Feynman-type propagator. 
Since the tensor $B^{m}{}_{ijm}$ is quadratic in the entries of the  matrix $M_{ij}$, it is a homogeneous  fourth order polynomial in the components of $q_i$. Also $\lambda$ is a homogeneous  fourth order polynomial in $q_i$. Consequently the propagator   ${}^F\!D_{ij}\sim q^{-2}$ 
%is inverse quadratic  in the wave covector  $q_i$ 
as the ordinary non-modified photon propagator. From (\ref{prop2add9}), we conclude that  the Landau-type propagator has the same behavior ${}^L\!D_{ij}\sim q^{-2}$.

%----------------------------------------
\section{Propagator for  skewon media}
In this section, we  calculate the propagator (\ref{prop11}) for a special case of a linear response medium -- the Minkowski vacuum modified by a generic skewon field. 
\subsection{Optic tensors and skewon covector}
The general constitutive tensor satisfies the symmetry relations
\begin{equation}\label{intr5}
 \chi^{ijmn}= -\chi^{jimn}=- \chi^{ijnm}\,,
\end{equation}
and thus has 36 independent components. Under the action of the general linear group $GL(4,\mathbb R)$, this  tensor is uniquely irreducibly decomposed into the  sum of three independent pieces \cite{birkbook},  
\begin{equation}\label{intr6}
\chi^{ijkl}={}^{\tt (1)}\!\chi^{ijkl}+{}^{\tt (2)}\!\chi^{ijkl}+ {}^{\tt(3)}\!\chi^{ijkl}\,.
\end{equation}

The {\it principal part} of 20 independent components  is expressed as
%\begin{eqnarray}\label{lin-7}  
%{}^{\tt (1)}\chi^{ijkl}& =&\frac %16\Big(2\left(\chi^{ijkl}+\chi^{klij}\right)
%-\left(\chi^{iklj}+\chi^{ljik}\right)-
%\nonumber\\&&\qquad
%\left(\chi^{iljk}+\chi^{jkil}\right)\Big)\,.
%\end{eqnarray}
\begin{equation}\label{lin-7}  
{}^{\tt (1)}\!\chi^{ijkl} =\frac 16\Big(2\chi^{ijkl}+2\chi^{klij}-\chi^{iklj}-\chi^{ljik}-\chi^{iljk}-\chi^{jkil}\Big)\,.
\end{equation}
It is a generalization of the standard Minkowski factor 
\begin{equation}\label{intr8}
{}^{\tt (1)}\!\chi^{ijkl}_M=g^{ik}g^{jl}-g^{il}g^{jk}\,.
\end{equation}
appearing in (\ref{intr2}). 
Two additional parts in (\ref{intr6}) do not have classical analogs. 

The {\it axion part}  ${}^{\tt(3)}\chi^{ijkl}$ is represented by  the complete antisymmetrization of $\chi^{ijkl}$ in all its four indices.  This pseudotensor has only one independent component. Consequently, it is  represented by a pseudoscalar $\alpha$,
\begin{equation}\label{intr9}
{}^{\tt (3)}\!\chi^{ijkl} =\chi^{[ijkl]}=\alpha\varepsilon^{ijkl}\,.
 \end{equation}
It is a rather popular object in modified electrodynamics.  

The {\it skewon part} ${}^{\tt (2)}\chi^{ijkl}$ of 15 independent components  is expressed as 
\begin{equation}\label{intr10}  
{}^{\tt (2)}\!\chi^{ijkl}=\frac
12\left(\chi^{ijkl}-\chi^{klij}\right)\,.
 \end{equation}
It is a primary object of our interest in the current paper. 
In order to clarify the skewon contribution to  electrodynamics, we will assume the principal part to be of the classical Minkowski form (\ref{intr8}). Moreover, we assume the flat Minkowski metric $g_{ij}={\rm diag}(-1,+1,+1,+1)$. Note that in a more general case of electrodynamics on a curved pseudo-Riemann manifold, ${}^{\tt (1)}\!\chi^{ijkl}$ must be endowed with a factor $\sqrt{-g}$. In our case, this factor is equal to 1 and thus omitted. 

With the constitutive tensor decomposed into the three independent pieces (\ref{intr6}), we obtain the optic tensor decomposed into two parts
\begin{equation}\label{skewprop1} 
M^{ij}=P^{ij}+Q^{ij}\,.
 \end{equation}
Here the principal part of the optic tensor is given by 
\begin{equation}\label{skewprop2} 
P^{ij}={}^{(1)}\!\chi^{imjn}q_mq_n\,,
 \end{equation}
while the skewon part  is 
\begin{equation}\label{skewprop3} 
Q^{ij}={}^{(2)}\!\chi^{imjn}q_mq_n\,.
 \end{equation}
The axion part drops out from the optic tensor. Consequently, in the framework of geometric optics approximation, axion does not contribute to the dispersion relation and to the photon propagator. 

The contributions of the principal and skewon parts  to the optic tensor  are of  very special forms. In particular, the principal part builds up the symmetric optic tensor while the skewon  optic tensor is skew-symmetric
\begin{equation}\label{skewprop4} 
P^{ij}=P^{ji}\,, \qquad Q^{ij}=-Q^{ji}\,,
 \end{equation}
Moreover, both these matrices are singular. Indeed, we evidently have the identities 
\begin{equation}\label{skewprop5} 
P^{ij}q_i=0\,, \qquad Q^{ij}q_i=0\,,
 \end{equation}
that mean  linear relations between the rows and the columns of the corresponding matrices. 

For  the antisymmetric tensor $Q^{ij}$, we can derive from (\ref{skewprop5}) a simple  representation, see \cite{Itin:2014gba}. Indeed, the expression  
\begin{equation}\label{skewprop6} 
 Q^{ij}=\ep^{ijkm}q_jY_m\,,
 \end{equation}
with an arbitrary covector $Y_m$ satisfies the second equation of (\ref{skewprop5}). For a given tensor $Q_{ij}$, the skewon covector $Y_m$ is determined from (\ref{skewprop6}) only up to an 
addition of a covector proportional to the wave covector, $Y_m\to Y_m+Cq_m$. This is an independent gauge invariant property, that  can be used for simplification of the calculations, see \cite{Itin:2013ica}. 

We consider the  special Minkowski principal part (\ref{intr8}).  In this case, Eq. (\ref{skewprop2}) yields
\begin{equation}\label{skewprop7} 
P^{ij}= g^{ij}q^2-q^iq^j\,.
 \end{equation}
The skewon part, however, is left to be of the generic form.
%%%%%%%
%%%%%%%%%%
\subsection{First adjoint}
In this sub-section  we present the first adjoint of the tensor $M^{ij}$ with the Minkowski symmetric part given in Eq.(\ref{skewprop7}) and the skewon antisymmetric part represented in Eq.(\ref{skewprop6}).  

It is convenient to start with a more general case  with   arbitrary principal  and  skewon parts. In this case,  the first adjoint tensor can be expressed as,
\begin{equation}\label{first1} 
A_{ij}=\left(\lambda_0+P^{mn}Y_mY_n\right)q_iq_j\,,
 \end{equation}
 see  \cite{Itin:2014gba} for details and proofs. 
Here the term $\lambda_0q_iq_j$ is the adjoint of the principal part matrix $P^{ij}$ alone. 
Consequently, the $\lambda$-term for the skewon modified media is given by 
\begin{equation}\label{first2} 
\lambda=\lambda_0+P^{mn}Y_mY_n\,.
 \end{equation}
For  source-free regions with $j^i=0$, a non-trivial  solution  (electromagnetic wave) exists if and only if the adjoint is equal to zero. This condition represents the {\it dispersion relation} that in our case takes the form
\begin{equation}\label{first2a} 
\lambda_0+P^{mn}Y_mY_n=0\,.
 \end{equation}
For the Minkowski principale part (\ref{intr8}) with the tensor $P^{ij}$ given in Eq.(\ref{skewprop7}), the  $\lambda$-term (\ref{first2}) takes a simple form \cite{Itin:2014gba}
\begin{equation}\label{first3} 
\lambda=q^4-Y^2q^2+(Y,q)^2\,.
 \end{equation}
The dispersion relation, in this case, is represented by a 4-th order homogeneous  equation 
\begin{equation}\label{first4} 
q^4-Y^2q^2+(Y,q)^2=0\,.
 \end{equation}

\subsection{Second adjoint}
We calculate now the second adjoint of the optic tensor  $M^{ij}$. 
Substituting the decomposition of the constitutive tensor (\ref{skewprop1}) into the expression  for the second adjoint 
 \begin{equation}\label{sec1} 
B_{ijkl}=\frac 1{2!}  \ep_{ijab}\ep_{klcd}M^{ac}M^{bd}\,.
 \end{equation}
we obtain it decomposed into a sum of three pieces
\begin{equation}\label{skewprop8}  
B_{ijkl}={}^{(1)}B_{ijkl}+{}^{(2)}B_{ijkl}+{}^{(3)}B_{ijkl}\,,
 \end{equation}
Here, ${}^{(1)}B_{ijkl}$ denotes  the $P^2$-term  expressed as
\begin{equation}\label{skewprop9}  
{}^{(1)}B_{ijkl}=\frac 1{2!}\ep_{ijmn}\ep_{klrs}P^{mr}
P^{ns}\,,
 \end{equation}
Substituting here the expression (\ref{skewprop7}) we calculate, see Appendix B,
\begin{equation}\label{skewprop9a}
{}^{(1)}B_{ijkl}=
q^2 \left(q_iq_kg_{jl}-q_jq_kg_{il}+q_jq_lg_{ik}-q_iq_lg_{jk}\right)\,.
 \end{equation}
The contraction of this expression reads
\begin{equation}\label{skewprop9b}
{}^{(1)}B_{ijkl}g^{jk}=
-\left(q^2g_{il}+2q_iq_l\right)q^2\,.
 \end{equation}
This is an expression related to the standard electrodynamics. 
The term ${}^{(2)}B_{ijkl}$ represents the mixed $PQ$-contribution 
\begin{equation}\label{skewprop10}  
{}^{(2)}B_{ijkl}= \ep_{ijmn}\ep_{klrs}P^{mr}Q^{ns}\,.
 \end{equation}
Calculations provided in Appendix B give
\begin{equation}\label{skewprop10a}  
{}^{(2)}B_{ijkl}=\left( \ep_{ijmk}q_l- \ep_{ijml}q_k\right) \left(Y^mq^2-q^m(q,Y) \right)\,.
\end{equation}
The contraction of this expression yields
\begin{equation}\label{skewprop10b}  
{}^{(2)}B_{ijkl}g^{jk}=- \ep_{ijml} Y^mq^jq^2\,.
\end{equation}
The third term ${}^{(3)}B_{ijkl}$ represents the pure skewon contribution
\begin{equation}\label{skewprop11}  
{}^{(3)}B_{ijkl}=\frac 1{2!}\ep_{ijmn}\ep_{klrs}Q^{mr}
Q^{ns}\,,
 \end{equation}
Its computation in Appendix B yields
\begin{equation}\label{skewprop11a}  
{}^{(3)}B_{ijkl}=(q_iY_j-q_jY_i)(q_kY_l-q_lY_k)\,,
 \end{equation}
The contraction of this expression reads
\begin{equation}\label{skewprop11b}  
{}^{(3)}B_{ijkl}g^{jk}=(q,Y)(q_iY_l+q_lY_i)-q^2Y_iY_l-Y^2q_iq_l\,.
 \end{equation}
%The Lorentz-type gauge condition $(q,Y)=0$ removes the first term, thus we are left with
%\begin{equation}\label{skewprop11c}  
%{}^{(3)}B_{ijkl}g^{jk}=-q^2Y_iY_l-Y^2q_iq_l\,.
% \end{equation}
\subsection{Propagator}
Substituting (\ref{skewprop9b},\ref{skewprop10b}) and (\ref{skewprop11b}) into the formula (\ref{prop11}) we obtain the propagator tensor
\begin{eqnarray}\label{phoprop1} 
D_{ij}&=&-\frac 1{\lambda q^2} \, \Big(
\left(q^2g_{ij}+2q_iq_j\right)q^2
+\ep_{ikmj} q^kY^mq^2 +\nonumber\\&&
q^2Y_iY_j+Y^2q_iq_j-(q,Y)(q_iY_j+q_lY_j)
\Big)\,.
 \end{eqnarray}
%We recall that this expression yields in the skewon gauge $(q,Y)=0$.  
We can simplify this expression by applying the gauge invariance of the propagator. 
In order to have the Feynman propagator, we merely remove from this expression all the terms proportional to the components $q_i$ and $q_j$. 
Consequently, the $q^2$ factor in the denominator cancels out and we are left with the following compact expression for the Feynman propagator
\begin{equation}\label{phoprop2} 
{}^F\!D_{ij}=-\frac 1{\lambda} \, \Big(
g_{ij}q^2+\ep_{ikmj} q^kY^m+Y_iY_j\Big)\,.
 \end{equation}
Explicitly,  it reads
\begin{equation}\label{phoprop3} 
{}^F\!D_{ij}=-\frac {g_{ij}q^2+\ep_{ikmj} q^kY^m+Y_iY_j} 
{q^4-q^2Y^2+(q,Y)^2} \,.
 \end{equation}
The Landau propagator also takes a simple form
\begin{equation}\label{phoprop5} 
{}^L\!D_{ij}=-\frac {\left(g_{ij}q^2-q_iq_j\right)+\ep_{ikmj} q^kY^m+Y_iY_j} {q^4-q^2Y^2+(q,Y)^2} \,.
 \end{equation}
Observe some basic properties of these propagators:
\begin{itemize}
\item[(1)] For the vanishing skewon field, the propagators return to their standard vacuum form
\begin{equation}\label{phoprop6} 
{}^F\!D_{ij}=-\frac {g_{ij}} {q^2}\,,\qquad {}^L\!D_{ij}=-\frac {g_{ij}q^2-q_iq_j} {q^4} \,.
 \end{equation}
The same is true also for the non-zero but physically trivial skewon covector $Y_i\sim q_i$.
\item[(2)] Skewon propagator has  nontrivial  symmetric and  antisymmetric parts. In the case of the Feynman gauge, they are expressed as 
\begin{equation}\label{phoprop7a} 
{}^F\!D_{(ij)}=-\frac {g_{ij}q^2+Y_iY_j} 
{q^4-q^2Y^2+(q,Y)^2}\,, 
 \end{equation}
and
\begin{equation}\label{phoprop7b} 
 {}^F\!D_{[ij]}=-\frac {\ep_{ikmj} q^kY^m} {q^4-q^2Y^2+(q,Y)^2}\,,
 \end{equation}
respectively. 
\item[(3)] The propagator is singular in at most four roots of the denominator.  As a result, there are at most two light cones at every space-time point -- birefringence.
\item[(4)] If the skewon covector $Y_i$ is regular in $q$ then in the first order approximation for a small skewon field
\begin{equation}\label{phoprop8} 
{}^F\!D_{(ij)}=-\frac {g_{ij}} 
{q^2}\,, \qquad {}^F\!D_{[ij]}=-\frac {\ep_{ikmj} q^kY^m} {q^4}\,.
 \end{equation}
Thus, the small skewon field affects only the antisymmetric part of the propagator. 
\end{itemize}
\section{Skewon and modified Coulomb law}
Since skewon modifies the propagator expression, it is naturally to expect the corresponding modification of the basic electrodynamics relationships, in particular, the Coulomb law. In order to describe such modification, we need an expression of propagator in the Coulomb gauge. 
\subsection{Skewon propagator in the Coulomb gauge}
Since we are dealing with the static problem, we need only one component of the propagator in the Coulomb gauge, namely ${}^C\!{D}_{00}$. From Eq.   (\ref{phoprop2}), we extract the following symmetric components:
\begin{equation}\label{coul-1}
{}^F\!{D}_{00}=\frac{q^2-Y_0^2}{\lambda}\,,\qquad {}^F\!\check{D}_{0\a}=-\frac{Y_0Y_\a}{\lambda}\,,
\end{equation}
and 
\begin{equation}\label{coul-2}
{}^F\!\check{D}_{\a\b}=-\frac{q^2g_{\a\b}+Y_\a Y_\b}{\lambda}\,.
\end{equation}
With these expressions, we calculate the temporal part (\ref{col9}) of the skewon propagator in the Coulomb gauge.  It reads
\begin{equation}\label{coul-3}
{}^C\!{D}_{00} =\frac{q^4k^2-\left(k^2Y_0-\omega(k,Y)\right)^2}{\lambda k^4}\,,
\end{equation}
where we use the 3-dimensional scalar product $(k,Y)=g^{\a\b}k_\a Y_\b$. 
Applying the Lorentz gauge for the skewon covector, $(Y,q)=0$, we replace the scalar product $(k,Y)$ by $\omega Y_0$ and rewrite this expression in the form
\begin{equation}\label{coul-4}
{}^C\!{D}_{00}=\frac{q^4(k^2-Y^2_0)}{\lambda k^4}\,.
\end{equation}
With the expression of the $\lambda$-function listed in Eq.(\ref{first3}), we have here 
\begin{equation}\label{coul-5}
{}^C\!{D}_{00}=\frac{q^2(k^2-Y^2_0)}{ k^4(q^2-Y^2+(q,Y)^2)}\,.
\end{equation}
For zero skewon, we are left here with the standard expresion ${}^C\!{D}_{00}={1}/{ k^2}$. 
 \subsection{Antisymmetric skewon}
We consider now some explicit model of the skewon field. 
As it is shown in \cite{birkbook}, a generic  skewon  $^{(2)}\chi^{ijkl}$ of 15 independent components can be uniquely represented by a traceless matrix $S_i{}^j$.  Since we are dealing with the   space endowed with a metric tensor $g_{ij}$, the mixed   tensor  $S_i{}^j$ can be replaced by the covariant and contravariant tensors
\begin{equation}\label{es-0}
S^{ij}=g_{ik}S_k{}^j\,,\quad{\rm and}\quad 
S_{ij}=g_{jk}S_i{}^k \,.
\end{equation}
These two  tensors depend not only of the skewon itself, but also of the metric tensor. In this paper, we restrict ourselves to the Minkowski metric, so this fact is irrelevant.  The tensors (\ref{es-0}) have some advantage because  they  can be invariantly decomposed into   symmetric and skew-symmetric parts. These two parts do not mix under coordinate transformations, those  they can be studied separately. 

Up to an addition of an arbitrary term proportional to the wave covector $q_i$, the skewon covector $Y_i$ can be presented in the form 
\begin{equation}\label{es-4}
Y_i=S_{ij}q^j\,.
\end{equation}

 The case of an antisymmetric matrix $S_{ij}=-S_{ji}$ is especially simple  because it provides the skewon covector of the regular polynomial form (\ref{es-4}) even for the Lorentz-type gauge, $(Y,q)=0$. We use the parametrization given in \cite{Itin:2014gba}. Moreover, we consider a special 3-parametric skewon of the "electric" type
\begin{equation}\label{as0} 
S_{0\mu}=-S_{\mu 0}=\a_\mu\,.
 \end{equation}
Correspondingly, 
\begin{equation}\label{as1} 
Y_0=\a_\mu k^\mu\,,\qquad Y_\mu=-\a_\mu\omega\,.
 \end{equation}
We use here the wave covector parametrized as $q^i=(\omega, k^\mu)$.  Observe that the skewon (\ref{as1}) indeed satisfies the gauge condition $(Y,q)=0$. 

We calculate the square of the skewon covector 
\begin{equation}\label{as1a} 
Y^2=\a^2\omega^2-(\a,k)^2\,.
 \end{equation}
Consequently, the denominator of the propagator takes the form
\begin{equation}\label{as2} 
\lambda=(\omega^2-{ k}^2) \left((1+\a^2)\omega^2-(\a,{ k})^2-{ k}^2\right)\,.
 \end{equation}
The zero set of this function has two branches -- two light cones. 
In addition to the ordinary circular light cone 
\begin{equation}\label{as3}
\omega^2={k}^2\,,
 \end{equation}
 there is an elliptic light cone 
\begin{equation}\label{as4}
(1+\a^2)\omega^2=(\a,{ k})^2+{ k}^2\,.
 \end{equation}
Comparing the expressions (\ref{as3}) and (\ref{as4}) we read off the following consequences, see \cite{Itin:2014gba} for detailed discussion:
 \begin{itemize}
\item[(1)] For an arbitrary real non-zero vector  $\a_\mu$, there are two separated cones. In optics, this behavior is refereed to as birefringence. 
\item[(2)] The elliptic cone is exterior to the circular one. 
It means superluminal wave propagation along it. In other words, the causality is broken down in this model for an arbitrary non-zero vector  $\a_\mu$.
\item[(3)] The cones meet one another along two  straight lines -- optic axes.
\end{itemize}

%--------------------
\subsection{Modification of the Coulomb law}
In order to study how the skewon field modifies the Coulomb law, we use the skewon propagator in the Coulomb gauge. Substituting (\ref{as1} -- \ref{as1a}) into (\ref{coul-5}), we obtain
\begin{equation}\label{as5}
{}^C\!{D}_{00}=\frac{(-\omega^2+k^2)(k^2-(\a,k)^2)}{ k^4\big[-\omega^2(1+\a^2)+k^2+(\a,k)^2\big]}\,.
 \end{equation}
In the non-relativistic limit, i.e., neglecting  the radiation effects,  we can neglect the term $\omega^2/c^2$ relative to $k^2$. Consequently we are left with 
\begin{equation}\label{as6}
{}^C\!{D}_{00}=\frac 1{k^2} \, \frac{k^2-(\a,k)^2}{ k^2+(\a,k)^2}\,.
 \end{equation}
Due to the experimental bounds provided recently by Ni, see  \cite{Ni:2013uwa},\cite{Ni:2014cca}, the skewon effect must be  very small.  Thus we use the approximation $(\a,k)^2<< k^2$. Consequently we obtain the temporal part of the skewon modified photon propagator in the form
\begin{equation}\label{as7}
{}^C\!{D}_{00}=\frac 1{k^2} \left(1-2 \frac{(\a,k)^2}{ k^2}\right)\,.
 \end{equation}
The electromagnetic potential for a point-wise charge $q$ is calculated by the inverse Fourier transform
\begin{equation}\label{as8}
 \vp(r)=q\int {}^C\!{D}_{00} e^{i(r,k)} \frac{d^3k}{(2\pi)^3}\,.
 \end{equation}
The first term of (\ref{as7}) provides the standard Coulomb potential  $\vp={q}/{(4\pi r)}$.
The second term presents the skewon addition, that  is given by the integral
\begin{equation}\label{as9}
\delta \vp=-2q\int\frac{(\a,k)^2}{ k^4}e^{i(r,k)} \frac{d^3k}{(2\pi)^3}\,.
 \end{equation}
This integral expression is familiar from the Breit equation in QED, see \cite{Landau}. 
We have the result of the integration 
\begin{equation}\label{as10}
\delta \vp=-\frac{q}{4\pi r}\left(\a^2-\frac{(\a,r)^2}{r^2}\right)\,.
 \end{equation}
Consequently, the skewon addition provides anisotropic modifications of the Coulomb potential 
\begin{equation}\label{as11}
 \vp=\frac{q}{4\pi r}\left(\left(1-\a^2\right)+\frac{(\a,r)^2}{r^2}\right)\,.
 \end{equation}
 The corresponding equipotential surfaces are ellipsoids.  They are similar to the wave-front surfaces of the skewon model \cite{Itin:2014rwa}.

\section{Results and discussion}
For a generic skewon part of the constitutive tensor, we evaluated the modified photon propagator in Feynman, Landau and Coulomb gauges. This tensor is asymmetric and singular on the Fresnel hypersurface. It is quite naturally to guess that the same features will emerge for a more generic medium with an arbitrary linear response tensor. In the lowest  order approximation, we derived the anisotropic modification of the Coulomb law. 

Some natural  problems  arise  in this context:
\begin{itemize}
\item What physical interpretation can be given to the antisymmetric part of the propagator?
\item Which new types of the modification of the Coulomb law emerge from the higher order approximations?  
\item How the two branches of the birefringent Fresnel hypersurface influence the modified  Coulomb law?
\item Which new observational effects  such modification can provide for the atom spectrum? These effects must be similar to the recently calculated splitting of the energy levels  originated in the Finsler-type modifications of the Coulomb law \cite{Itin:2014uia}. 

\end {itemize}
\begin{acknowledgments}
The author would like to thank  F.-W. Hehl (Cologne/ Missouri), Wei-Tou Ni (Hsinchu, Taiwan), Yu. Obukhov (Moskow),  V. Perlick (ZARM, Bremen), and Y. Friedman (JCT, Jerusalem)  for valuable comments.  
\end{acknowledgments}
\appendix
\section{Determinant and adjoints}
\subsection{Determinant of a tensor}
In four-dimensional space, the determinant of a covariant (up-indexed) second order tensor $M^{ij}$ can be written as 
\begin{equation}\label{det1} 
{\rm det} \, M=\frac 1{4!}\ep_{ijkl}\ep_{prst}
M^{ip}M^{jr}M^{ks}M^{lt}\,.
 \end{equation}
The Levi-Civita symbol takes the values  $\epsilon_{ijkl}=\{+1,-1,0\}$, for even, odd, and no permutation of the indices $\{0123\}$, respectively; the analogous is assumed for the up-indexed Levi-Civita's  symbol $\epsilon^{ijkl}$.
It is can be  checked straightforwardly  that the expression on the right-hand side of Eq.(\ref{det1}) includes all proper  products of the elements of the matrix $M^{ij}$ with the  proper signs that usually appear in the determinant.
The  leading factor $1/{4!}$ provides the correct value of the determinant, thus the formula  in Eq.(\ref{det1}) is proved. 

%%%%%%%%%%%%%%%%%%
\subsection{First adjoint tensor}
The cofactor matrix of a square matrix  is constructed by removing one row and one column of the original matrix and calculating the determinants of the remaining  matrices. The transpose of the cofactor matrix is called (classical)  adjoint or adjugate, or adjunct matrix. In order to simplify the formulas above we use the adjoint without transpose.

For a covariant (up-indexed) $4\times 4$ tensor  $M^{ij}$ we define the {\it adjoint tensor} $A_{ij}={\rm adj} (M)$ to be expressed by a  contravariant (down-indexed) second order tensor
\begin{equation}\label{det2} 
A_{ij} =\frac 1{3!}\ep_{iabc}\ep_{jdef}
M^{ad}M^{be}M^{cf}\,.
 \end{equation}
%It is easy to recognize  
For some applications, it is convenient to see the adjoint tensor as a derivative of the determinant relative to the tensor entries
\begin{equation}\label{det2x} 
A_{ij} =\frac{\partial\,{\rm det} M}{\partial M^{ij}}\,.
 \end{equation}

The classical Laplace expansion of the determinant is written in tenor notations as
\begin{equation}\label{det3} 
M^{ij}A_{kj} =\left({\,\rm det} M\right)\d^i_k\,,
 \end{equation}
or,
\begin{equation}\label{det3a} 
M^{ij}A_{ik} =\left({\,\rm det} M\right)\d^j_k\,.
 \end{equation}
Here the first equation represents the expansion of a determinant along its rows  while the second equation means the expansion along the columns. This equation follows straightforwardly from the definitions (\ref{det1}--\ref{det2}). 
Using the transpose of $A_{kj}$ we return in Eq.(\ref{det3}) to the classical expansion of the determinant with the standard "row-times-column" rule.

In the non-singular case, ${\rm det} (M)\ne 0$, tensor $M^{ij}$ has an inverse tensor that is necessary down-indexed. It is expressed as 
\begin{equation}\label{det3x} 
\left(M^{-1}\right)_{ij}=\frac{1}{{\,\rm det} M}\,A_{ij}\,.
 \end{equation}

\subsection{Second adjoint}

The {\it second adjoint} is defined by removing {\it two rows and two columns} from the original matrix. Formally it can be  written as 
 \begin{equation}\label{det7} 
B_{ijkl}=\frac 1{2!}  \ep_{ijab}\ep_{klcd}M^{ac}M^{bd}\,.
 \end{equation}
From this expression, we read off the symmetries
 \begin{equation}\label{prop4} 
B_{ijkl}=-B_{jikl}=-B_{ijlk}\,.
 \end{equation}
We can also arrange this tensor as the derivative of the first adjoint tensor and even of the determinant itself
\begin{equation}\label{det8} 
B_{ijkl} =\frac{\partial\,A_{ik}}{\partial M^{jl}}=\frac{\partial^2{\rm det}\,M}{\partial M^{ik}\partial M^{jl}}\,.
 \end{equation}

The  Laplace identity for the second adjoint  is readily derived from the identity given in Eq.(\ref{det3}). Taking a derivative relative to the entries of the tensor $M^{rs}$ we obtain
 \begin{equation}\label{ap-prop5} 
\frac{\partial M^{ij}}{\partial M^{rs}}A_{kj}+M^{ij}\frac{\partial A_{kj}}{\partial M^{rs}}=\d^i_k\frac{\partial \,{\rm det}\,M}{\partial M^{rs}}\,.
 \end{equation}
We use the differential identities (\ref{det2x}) and (\ref{det8}) to obtain 
\begin{equation}\label{prop6a} 
\d^i_r\d^j_sA_{kj} +M^{ij}B_{krjs}=\d^i_kA_{rs}\,.
 \end{equation}
Consequently, the  Laplace identity reads
 \begin{equation}\label{ap-prop6} 
M^{ij}B_{krjs}=\d^i_kA_{rs}-\d^i_rA_{ks}\,.
 \end{equation}
We observe some consequences of this identity. Contracting the indices  $i$ and $s$, we get
 \begin{equation}\label{ap-prop7} 
M^{ij}B_{krji}=A_{rk}-A_{kr}\,.
 \end{equation}
This is in a correspondence with the symmetries (\ref{prop4}). Contracting of (\ref{prop6}) relative to the indices $i$ and $k$, we have
 \begin{equation}\label{ap-prop7a} 
M^{ij}B_{irjs}=3A_{rs}\,.
 \end{equation}
The latter identity is a consequence of the order 3 homogeneity of the adjoint tensor. 

Using the same derivative procedure for the identity (\ref{det3a}) we derive the second Laplace identity 
 \begin{equation}\label{ap-prop8} 
M^{ij}B_{irks}=\d^j_kA_{rs}-\d^j_sA_{rk}\,.
 \end{equation}

\section{Calculation of the second adjoint}
In this section, we provide the details of the calculations of the second adjoin (\ref{skewprop9}).

%%%%%
\subsection{Calculation of ${}^{(1)}B_{ijkl}$}%{skewprop9}
Substituting  (\ref{skewprop7}) into the first part of  (\ref{skewprop9}) we have
\begin{eqnarray}\label{apen-01}  
&&{}^{(1)}B_{ijkl}=\frac 1{2!}\ep_{ijmn}\ep_{klrs}P^{mr}
P^{ns}\nonumber\\
&=&\frac 1{2}\ep_{ijmn}\ep_{klrs}\left(g^{mr}q^2-q^mq^r\right)\left(g^{ns}q^2-q^nq^s\right)
\end{eqnarray}
Using the standard formulas for the contraction of two $\ep$-symbols, we obtain 
\begin{eqnarray}\label{apen-01a}  
{}^{(1)}B_{ijkl}&\!\!=\!\!& q^4\left| \begin{array}{cc} g_{ik}&g_{il}\\ g_{jk}&g_{jl} \end{array} \right|
-q^2q^mq^r
\left| \begin{array}{ccc}
g_{ik} &g_{il}&g_{ir} \\
g_{jk} &g_{jl}&g_{jr}  \\
g_{mk} &g_{ml}&g_{mr} 
\end{array} \right|\nonumber\\
&&
\end{eqnarray}
We multiply the last row and the last column of the second determinant by $q^r$ and $q^m$ correspondingly  
\begin{equation}\label{apen-01b}  
{}^{(1)}B_{ijkl}= q^4\left| \begin{array}{cc} g_{ik}&g_{il}\\ g_{jk}&g_{jl} \end{array} \right|
-q^2
\left| \begin{array}{ccc}
g_{ik} &g_{il}&q_{i} \\
g_{jk} &g_{jl}&q_{j}  \\
q_{k} &q_{l}&q^2
\end{array} \right|
\end{equation}
Evaluating the second determinant, we obtain 
\begin{equation}\label{apen-01c}  
{}^{(1)}B_{ijkl}=q^2 \left(q_j\left| \begin{array}{cc}
g_{ik} &g_{il}  \\
q_{k} &q_{l}
\end{array} \right|-q_i\left| \begin{array}{ccc}
g_{jk} &g_{jl}  \\
q_{k} &q_{l}
\end{array} \right|
\right)\,.
\end{equation}
Consequently, 
\begin{equation}\label{apen-01d}  
{}^{(1)}B_{ijkl}=
q^2 \left(q_iq_kg_{jl}-q_jq_kg_{il}+q_jq_lg_{ik}-q_iq_lg_{jk}\right)\,.
\end{equation}

\subsection{Calculation of ${}^{(2)}B_{ijkl}$}
Let us calculate the mixed term of the second adjoint 
\begin{eqnarray}\label{apen-02a}  
{}^{(2)}B_{ijkl}&\!\!=\!\!& \ep_{ijmn}\ep_{klrs}P^{mr}
Q^{ns}\nonumber\\
&\!\!=\!\!&\ep_{ijmn}\ep_{klrs}\left( q^2g^{mr}-q^mq^r\right)
\ep^{nspt}q_pY_t
 \end{eqnarray}
Evaluating the product of two $\ep$-symbols in the term of a determinant, we have
\begin{eqnarray}\label{apen-02b}  
{}^{(2)}B_{ijkl}&\!\!=\!\!& \ep_{ijmn}\left( q^2g^{mr}-q^mq^r\right)q_pY_t
\left| \begin{array}{ccc}
\d^{n}_{k} &\d^{p}_{k}&\d^{t}_{k} \\
\d^{n}_{l} &\d^{p}_{l}&\d^{t}_{l} \\
\d^{n}_{r} &\d^{r}_{r}&\d^{t}_{r} 
\end{array} \right|
\nonumber\\
&\!\!=\!\!&\ep_{ijmn}\left( q^2g^{mr}-q^mq^r\right)
\left| \begin{array}{ccc}
\d^{n}_{k} &q_{k}&Y_{k} \\
\d^{n}_{l} &q_{l}&Y_{l} \\
\d^{n}_{r} &q_{r}&Y_{r} 
\end{array} \right|\nonumber\\
&\!\!=\!\!&
 \left| \begin{array}{ccc}
\d^{n}_{k} &q_{k}&Y_{k} \\
\d^{n}_{l} &q_{l}&Y_{l} \\
0 &0&\ep_{ijmn}\left(Y^mq^2-q^m(q,Y) \right)
\end{array} \right|\nonumber\\
&\!\!=\!\!& \ep_{ijmn}\left( Y^mq^2-q^m(q,Y) \right)
\left| \begin{array}{cc}
\d^{n}_{k} &q_{k} \\
\d^{n}_{l} &q_{l}\\
\end{array} \right|%\nonumber\\\,,
 \end{eqnarray}
Consequently, we obtain
\begin{eqnarray}\label{apen-03}  
{}^{(2)}B_{ijkl}&=& \left( \ep_{ijmk}q_l- \ep_{ijml}q_k\right) \left(Y^mq^2-q^m(q,Y) \right)\,.
\end{eqnarray}

\subsection{Calculation of ${}^{(3)}B_{ijkl}$}

Let us calculate this expression in the term of the skewon optic covector. Substituting  (\ref{skewprop6}) into (\ref{skewprop11}), we  write it as 
\begin{eqnarray}\label{apen-2}  
{}^{(3)}B_{ijkl}&=&\frac 1{2!}  \ep_{iki_1i_2}\ep_{jlj_1j_2}Q^{i_1j_1}Q^{i_2j_2}\nonumber\\
&=&
\frac 1{2}\big(\ep_{ijmn}\ep^{mrab}q_aY_b\big)%\cdot 
\big(\ep_{klrs}\ep^{nscd}q_cY_d\big)\,.
\end{eqnarray}
Using the standard formulas for the product of two permutation tensors, we have
\begin{equation}\label{apen-3} 
 \ep_{ijmn}\ep^{mrab}q_aY_b=\left| \begin{array}{ccc}
\d^{r}_{i} &\d^{a}_{i}&\d^{b}_{i} \\
\d^{r}_{j} &\d^{a}_{j}&\d^{b}_{j} \\
\d^{r}_{n} &\d^{a}_{n}&\d^{b}_{n} 
\end{array} \right|q_aY_b=\left| \begin{array}{ccc}
\d^{r}_{i} &q_i&Y_i \\
\d^{r}_{j} &q_j&Y_j \\
\d^{r}_{n} &q_n&Y_n
\end{array} \right|\,.\end{equation}
Similarly,
\begin{equation}\label{apen-3x} 
\ep_{klrs}\ep^{nscd}q_cY_d=\left| \begin{array}{ccc}
\d^{n}_{k} &\d^{c}_{k}&\d^{d}_{k} \\
\d^{n}_{l} &\d^{c}_{l}&\d^{d}_{l} \\
\d^{n}_{r} &\d^{c}_{r}&\d^{d}_{r}
\end{array} \right|q_cY_d=\left| \begin{array}{ccc}
\d^{r}_{i} &q_k&Y_k \\
\d^{r}_{j} &q_l&Y_l \\
\d^{r}_{n} &q_r&Y_r
\end{array} \right|\,.\end{equation}
Thus
\begin{eqnarray}
{}^{(3)}B_{ijkl}&=&\frac 12 \left| \begin{array}{ccc}
\d^{r}_{i} &q_i&Y_i \\
\d^{r}_{j} &q_j&Y_j \\
\d^{r}_{n} &q_n&Y_n
\end{array} \right|\cdot
\left| \begin{array}{ccc}
\d^{r}_{i} &q_k&Y_k \\
\d^{r}_{j} &q_l&Y_l \\
\d^{r}_{n} &q_r&Y_r
\end{array} \right|%\nonumber
\end{eqnarray}
Expanding the third-order determinants, we have
\begin{eqnarray}
{}^{(3)}B_{ijkl}&=&
\frac 12 \left(\d^r_i\left| \begin{array}{cc}
q_j &Y_j \\
q_n &Y_n 
\end{array} \right|-
\d^r_j\left| \begin{array}{cc}
q_i &Y_i \\
q_n &Y_n 
\end{array} \right|+
\d^r_n\left| \begin{array}{cc}
q_i &Y_i \\
q_j &Y_j 
\end{array} \right|
\right)\nonumber\\&&\cdot
\left(\d^n_k\left| \begin{array}{cc}
q_l &Y_l \\
q_r &Y_r 
\end{array} \right|-
\d^n_l\left| \begin{array}{cc}
q_k &Y_k \\
q_r &Y_r 
\end{array} \right|+
\d^n_r\left| \begin{array}{cc}
q_k &Y_k \\
q_l &Y_l 
\end{array} \right|
\right)\nonumber\\
\end{eqnarray}
Term by term multiplication yields
\begin{eqnarray}
{}^{(3)}B_{ijkl}&=& 
\left| \begin{array}{cc}q_j &Y_j \\q_k &Y_k \end{array} \right|
\left| \begin{array}{cc}q_l &Y_l \\q_i &Y_i \end{array} \right|-
\left| \begin{array}{cc}q_j &Y_j \\q_l &Y_l \end{array} \right|
\left| \begin{array}{cc}q_k &Y_k \\q_i &Y_i \end{array} \right|\nonumber\\
&=&(q_jY_k-q_kY_j)(q_lY_i-q_iY_l)-\nonumber\\&&(q_jY_l-q_lY_j)(q_kY_i-q_iY_k)\nonumber\\
&=&(q_iY_j-q_jY_i)(q_kY_l-q_lY_k)\,.
\end{eqnarray}
Expanding these expressions, we come to a compact formula
\begin{eqnarray}
{}^{(3)}B_{ijkl}&=& (q_iY_j-q_jY_i)(q_kY_l-q_lY_k)\,.
\end{eqnarray}
It can be written also  in a matrix form
\begin{eqnarray}
{}^{(3)}B_{ijkl}&=&
\left| \begin{array}{cc}q_i &Y_i \\q_j &Y_j \end{array} \right|
\left| \begin{array}{cc}q_k &Y_k \\q_l &Y_l \end{array} \right|\,.
\end{eqnarray}

{}
\end{document}